\documentclass{aa}

\usepackage{graphicx,xspace}
\usepackage{natbib}
\usepackage{url}
\usepackage[varg]{txfonts}
\usepackage{threeparttable}
\def\ero{\textit{eROSITA}\xspace}

\def\gai{\textit{Gaia}\xspace}

\def\ros{\textit{ROSAT}\xspace}

\def\srg{\textit{SRG}\xspace}
\def\swi{\textit{Swift}/XRT\xspace}
\def\tes{\textit{TESS}\xspace}
\def\xmmn{\textit{XMM-Newton}\xspace}

\newcommand\srgt{{SRGt\,062340.2$-$265715}\xspace}

\newcommand\fergs{\ensuremath{\mathrm{erg}\,\mathrm{cm}^{-2}\,\mathrm{s}^{-1}}\xspace}

\newcommand\kmps{\ensuremath{\mathrm{km}\,\mathrm{s}^{-1}}\xspace}

\newcommand\rat{\ensuremath{\mathrm{s}^{-1}}\xspace}

\begin{document}

\title{Identification of SRGt\,062340.2-265715 as a bright, strongly variable, novalike cataclysmic variable}

\author{Axel Schwope\inst{1}
\and David A.H.~Buckley \inst{2, 3}
\and Adela Kawka\inst{4}
\and Ole König\inst{5}
\and Alexander Lutovinov\inst{6}
\and Chandreyee Maitra\inst{7}
\and Ilya Mereminskiy\inst{6}
\and James Miller-Jones\inst{4}
\and Manuel Pichardo Marcano\inst{8} 
\and Arne Rau\inst{7}
\and Andrei Semena\inst{6}
\and Lee J.~Townsend\inst{2}
\and Jörn Wilms\inst{5}
}
\institute{Leibniz-Institut f\"ur Astrophysik Potsdam (AIP), An der Sternwarte 16, 14482 Potsdam, Germany\\
\email{aschwope@aip.de}
\and
South African Astronomical Observatory, P.O. Box 9, Observatory, 7935, Cape Town, South Africa
\and
Department of Astronomy, University of Cape Town, Private Bag X3, Rondebosch 7701, South Africa
\and
International Centre for Radio Astronomy Research, Curtin University, GPO Box U1987, Perth, WA 6845, Australia
\and
Dr.~Karl Remeis-Sternwarte \& Erlangen Centre for Astroparticle Physics, Friedrich-Alexander-Universit\"at Erlangen-N\"urnberg, Sternwartstr.~7, 96049 Bamberg, Germany
\and
Space Research Institute (IKI) of Russian Academy of Sciences, Profsoyuznaya ul 84/32, 117997 Moscow, Russian Federation
\and 
     Max-Planck-Institut f\"ur extraterrestrische Physik,
     Gie{\ss}enbachstra{\ss}e, 85748 Garching, Germany
\and
Department of Physics and Astronomy, Texas Tech University, USA
}
\authorrunning{Schwope et al.}
\titlerunning{Identification of SRGt 062340.2-265715 as a bright cataclysmic variable}
\date{\today}

\keywords{stars: cataclysmic variables – X-rays: stars - stars: individual: SRGt 062340.2-265715}

\abstract
{We report the identification and follow-up of the transient \srgt detected with both instruments on board the Spektrum-Roentgen-Gamma mission. Optical spectroscopy of the $G=12.5$ counterpart firmly classifies the object as a novalike cataclysmic variable (CV) at a distance of 495\,pc. A highly significant \tes period of 3.941 hours, tentatively identified with the orbital period of the binary, could not be found when the object was reobserved with \tes two years later. The newer high-cadence \tes data revealed quasi-periodic oscillations around 25 min, while ground-based photometry indicated periodic variability at 32 min. Located in very sparsely populated regions of color-magnitude diagrams involving X-ray and optical magnitudes and colors, the new object could be an X-ray underluminous magnetic CV, an intermediate polar, or an overluminous nonmagnetic CV. The lack of uniquely identified spin and orbital periods prevents a final classification. The site of X-ray production in the system, $L_{\rm X, bol} = 4.8 \times 10^{32}$\,erg \rat, remains to be understood given its high variability on long and short timescales.
}

\maketitle

\section{Introduction}

In the course of the second all-sky survey, on 2020 October 12, the two instruments on board the Spektrum-Roentgen-Gamma mission \citep[{\it SRG},][]{sunyaev+21} detected a bright transient at galactic coordinates $l^{II} = 234\fd7, b^{II}=-17\fd6$. The comparison of data obtained during the first and second all-sky surveys with the two instruments on board the mission, the {\it Mikhail Pavlinsky} ART-XC \citep{2021arXiv210312479P} and \ero \citep{predehl+21}, revealed an increase in X-ray flux of about a factor 2, which triggered some fast-turnaround follow-up observations.

An initial announcement of the source was published by \citet{schwope+20} and some analysis of the \tes\ data was reported by \cite{pichardo20}. In this paper we describe the initial results in more detail, together with follow-up spectroscopy and photometry from the ground and from space. The transient was found to be coincident with cataloged \ros, \swi, and \gai sources (1RXS J062339.8$-$265744 and 2SXPS J062339.9$-$265751) and the optical transient ZTF19aaabzuh, respectively. It was mentioned earlier as a cataclysmic variable by Denisenko on his webpage\footnote{\url{http://scan.sai.msu.ru/~denis/VarDDE.html}}, but no further supporting information was available.

The X-ray and optical observations uniquely identify the transient with a nova-like cataclysmic variable (CV). CVs are close binaries with white dwarfs that accrete from a Roche-lobe filling star. In approximate terms, they may physically be divided into magnetic and nonmagnetic objects (disk accretion vs.~ quasi-radial accretion), and phenomenologically, they are often sorted into dwarf novae (DN, which are nonmagnetic disk-accreting CVs that may become unstable and cause the DN outburst) and so-called novalikes. Historically, both magnetic and nonmagnetic CVs were sorted into this last class. The nonmagnetic disk-accreting novalikes (often referred to as UX UMas, although the VY Scl, Z Cam, and SW Sex objects and possibly other subtypes may also fall in this class) are objects with high-accretion rates whose disks are permanently found in the hot state so that no dwarf nova outbursts are launched. CVs of all subtypes were expected to be found in large number through systematic spectroscopic follow-up of point-like sources serendipitously discovered in the ongoing all-sky surveys with \ero \citep{schwope12}. The case considered here is unusual through its very high optical brightness and its high and strongly variable X-ray brightness in the hard \srg band as well, which allowed its discovery with both instruments on the spacecraft.

\section{Observations and analysis\label{s:obs}}

\subsection{\srg}
The transient of 2020 October 12 was observed at the \ero position $\alpha_\mathrm{J2000.0}=06^\mathrm{h}23^\mathrm{m}40\fs2$ ($95\fdg91754$), $\delta_\mathrm{J2000.0}=-26^\circ57'15''$ ($-26\fdg96425$) with a positional uncertainty of $1\farcs6$ ($0\farcs6$ statistical, $1\farcs5$ systematic). The {\it Mikhail Pavlinsky} ART-XC telescope detected the source at the position of
$\alpha_\mathrm{J2000.0}=06^\mathrm{h}23^\mathrm{m}39\fs7$,
$\delta_\mathrm{J2000.0}=-26^\circ57'53''$
with an uncertainty of $15''$ (90\%).

During the second \srg\ survey, the object had a total exposure time of 85 s (an effective exposure time of 25\,s) with ART-XC and was discovered at a mean flux of $1.3_{-0.3}^{+0.4}\times10^{-11}$\,\fergs in the 4--12\,keV energy band. It was below the detection threshold during the first survey, when it had an effective exposure time of 21\,s, corresponding to an upper limit for the 4--12\,keV flux of $F < 10^{-11}$\,\fergs. With \ero, the source was detected in both surveys, in eRASS1 (2020 April 8), where it received a total exposure time of 228\,s, with a mean count rate of $1.38 \pm 0.11$\,\rat , and in eRASS2 (2020 October 12) with a mean rate of $5.84 \pm 0.24$\,\rat for a total exposure of 208\,s.

\begin{figure}
\resizebox{\hsize}{!}{\includegraphics{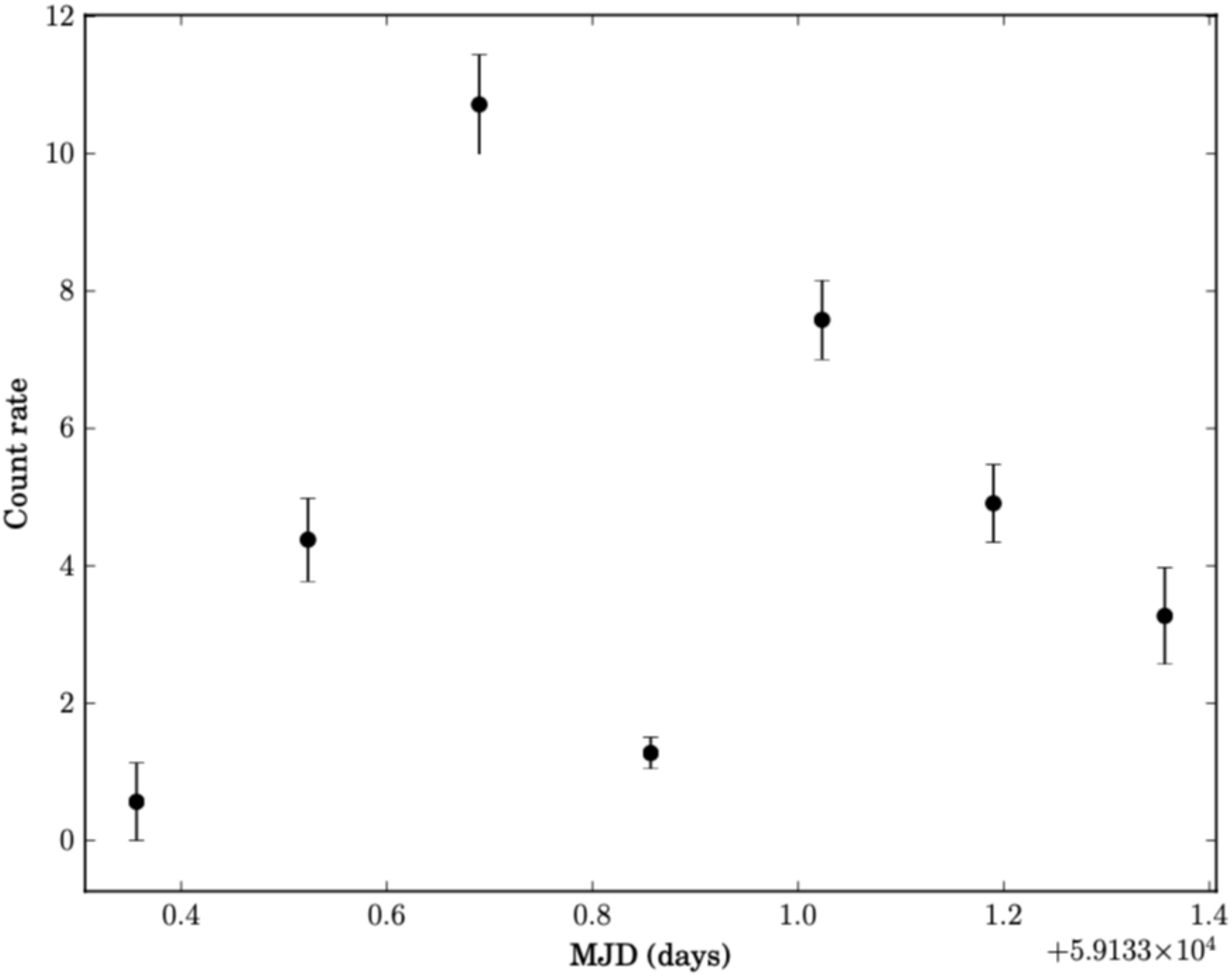}}
    \caption{X-ray light curve of \srgt obtained for eRASS2 (0.2 - 10 keV). Each data point represents the mean rate of one individual scan over the source position.}
    \label{f:elc}
\end{figure}

More than 1200 photons were collected during eRASS2, which allowed generating the eRASS2 light curve that is displayed in Fig.~\ref{f:elc}.  Circular regions with radii of 75\arcsec\, and 100\arcsec\, were used for the source and background extraction, respectively. \srgt was observed during seven erodays\footnote{An eroday lasts 4 hours and corresponds to the revolution of the spacecraft around the scan axis. The exposure for a given sky position is different for each eroday and may last up to 40 s dependent on the off-axis angle} and showed variability between $10.7 \pm 0.7$\,\rat and $1.3 \pm 0.2$\,\rat. Most of the photons registered with ART-XC were registered during the fourth and fifth eroday (see Fig.~\ref{f:elc}).

\begin{figure}
\resizebox{\hsize}{!}{\includegraphics{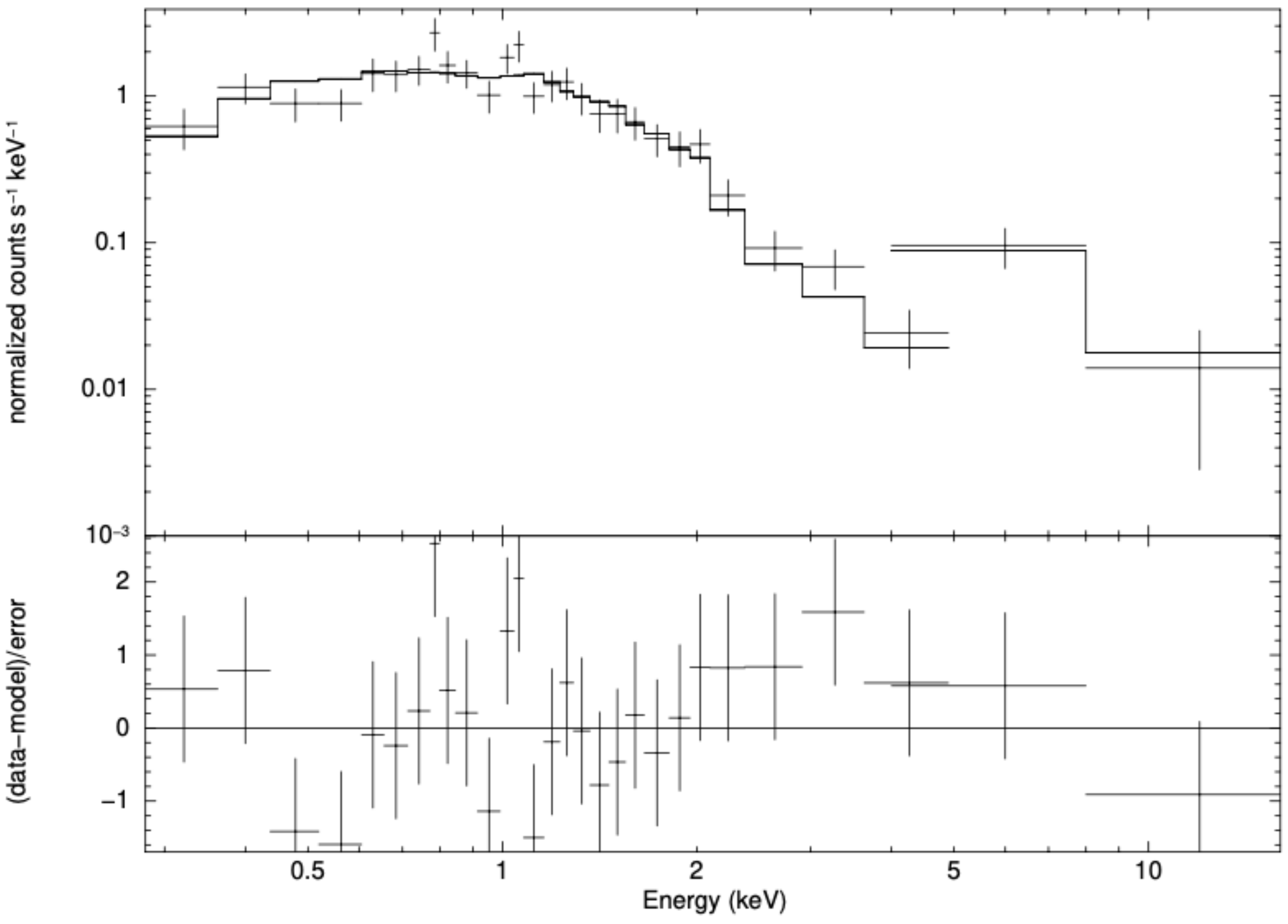}}
    \caption{Mean combined spectrum of ART X-C and \ero of \srgt obtained during the second survey with SRG with the best-fitting thermal model spectrum. The lower panel shows the residuals in units of the error per spectral bin.}
    \label{f:espc}
\end{figure}

The combined spectrum from the two instruments on board SRG is displayed in Fig.~\ref{f:espc}. It was fit using XSPEC version 12.11 with a one-component emission model, a thermal plasma absorbed by cold interstellar matter. A scaling factor between the two spectral groups was also fit that accounts (a) for calibration uncertainties between the two involved instruments and (b) for the fact that the data were not taken strictly simultaneously, which is important given the high variability of the source. The C-statistic was used for fitting. The best-fit value found was 180 for 230 bins. This provides a temperature $kT =5.7^{+2.6}_{-1.6}$\,keV (errors given for a 90\% confidence interval) and fluxes of 
$F_{\rm eRO}\, \mbox(0.5-2.0\,{\rm keV}) = 4.3^{+0.3}_{-0.4} \times 10^{-12}$ \,\fergs, 
$F_{\rm eRO}\, \mbox(0.5-10\,{\rm keV}) = 1.2^{+0.5}_{-0.4} \times 10^{-11}$ \,\fergs , and
$F_{\rm ART}\, \mbox(4-12\,{\rm keV}) = 2.0^{+0.14}_{-0.3} \times 10^{-11}$ \,\fergs.
The column density was determined using the {\tt{TBabs}} command (abundance set to {\it wilms}), but was not found to be well constrained. It was therefore fixed at the Galactic column density in this direction of N$_{\rm H} = 4.08 \times 10^{20}$\,cm$^{-2}$ \citep{h14pi16}. Recent 3D dust maps \citep[e.g.,][]{green+19} do not provide better constraints due to rather large errors, $E(g-r)=0.03\pm0.02$. The implied column density is compatible with that inferred from X-rays.
The same model applied to the eRASS1 data with a fixed amount of interstellar absorption gave a lower temperature $kT = 2.0^{+3.4}_{-0.6}$\,keV at a flux  of $F_{\rm X} \mbox(0.5-2.0\,{\rm keV}) = (8.4 \pm 1.6) \times 10^{-13}$\,\fergs. The bolometric flux of the best-fitting model during the high state in October 2020 was $F_{\rm X} = 1.6 \times 10^{-11}$\,\fergs (eROSITA value).

\begin{figure}
\resizebox{\hsize}{!}{\includegraphics{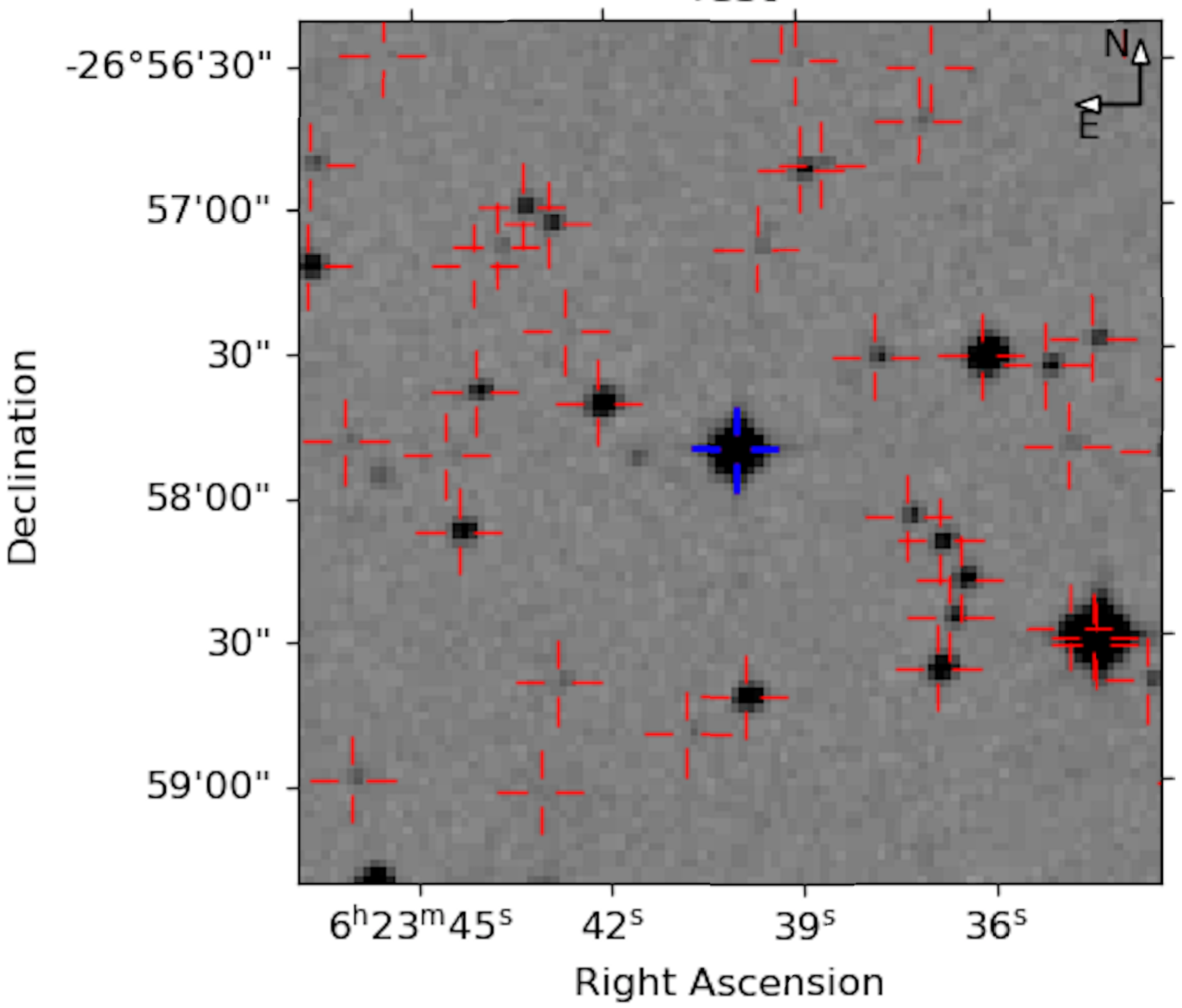}}
\caption{Finding chart of the new transient ($3\times3$ arcmin). The background image is from the DSS, and the eROSITA position (eRASS2) is marked with a blue cross. All further \gai objects in the field of view are marked with red crosses.
\label{f:fc}}
\end{figure}

\subsection{\gai}
\gai EDR3 lists an object (source\_ID 2899766827964264192) at a distance of $2\farcs4$ from the eRASS2 catalog position of the transient at position $\mathrm{RA} = 95\fdg91672681385$, $\mathrm{DEC}= -26\fdg96405593108$. A finding chart indicating the eRASS2 X-ray position and all \gai sources in its vicinity is reproduced in Fig.~\ref{f:fc}. It has a parallax of $\pi = 1.9749\pm0.0192\,\mathrm{mas}$ and photometric properties $G = 12.436 \pm0.005$\,mag and $Bp-Rp = 0.023$. Following \cite{bailer-jones+21}, the median geometric distance to \srgt  is $495.5\pm4$\,pc, which gives an absolute magnitude $M(G) = 3.96$. 

\begin{table*}
\caption{Archival X-ray observations of \srgt as derived from the ESA upper limit server HILIGT }
\label{t:xobs}
\begin{tabular}{l c r l l l}
\hline\hline
Mission & Date & Exp time & Flux$^{a}$ & Flux$^{a}$ & Flux$^{a}$ \\
 &  & (s) & 0.2-2 keV & 2-12 keV & 0.2-12 keV  \\ \hline
ROSAT-Survey & 1990-09-19 00:47:17 & 661.5 & $2.1 \pm 0.2$ & $$ & $$ \\
XMM-Newton slew & 2003-04-01 & 5.6 & $2.3 \pm 0.8$ & $<28.2$ & $6.4 \pm 2.0$ \\
XMM-Newton slew & 2006-10-02 & 9.9 & $5.0 \pm 0.8$ & $11.6 \pm 3.3$ & $12.4 \pm 1.8$ \\
Swift-XRT & 2008-07-17 & 5558.3 & $1.7 \pm 0.1$ & $2.3 \pm 0.2$ & $3.8 \pm 0.2$ \\
\hline
\end{tabular}

$^{a}$ Absorbed flux in units of $10^{-12}$\,\fergs. Count rate to flux conversion performed with a power-law spectral model of slope 2 and absorption $N_{\rm H} = 3\times10^{20}\,\mathrm{cm}^{-2}$.
\end{table*}

\subsection{Archival X-ray observations}

The position of \srgt was covered previously by past (\ros) and current (\xmmn\ and \swi) X-ray missions. The corresponding information is available via the ESA-supported high-energy light-curve generator HILIGT (upper limit server)\footnote{\url{http://xmmuls.esac.esa.int/hiligt/}}
and listed in Tab.~\ref{t:xobs}. Although HILIGT does not offer a thermal spectral model to convert count rates into fluxes, the given values are at least indicative.  The flux in the soft band, $0.2-2.0$\, keV, common to all the missions, was found to be variable by at least a factor 2.

\begin{figure}
\resizebox{\hsize}{!}{\includegraphics{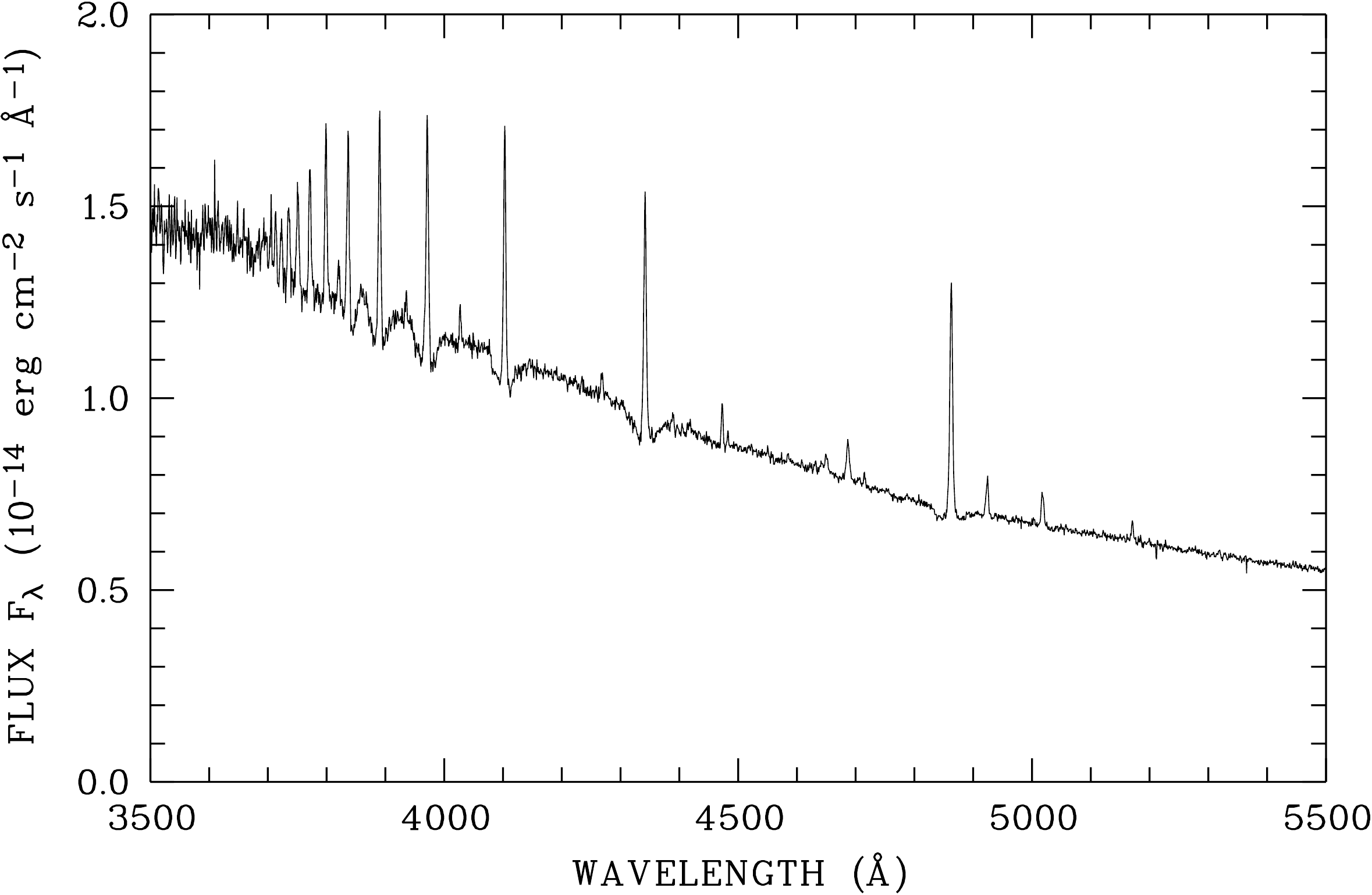}}
\resizebox{\hsize}{!}{\includegraphics[angle=-90]{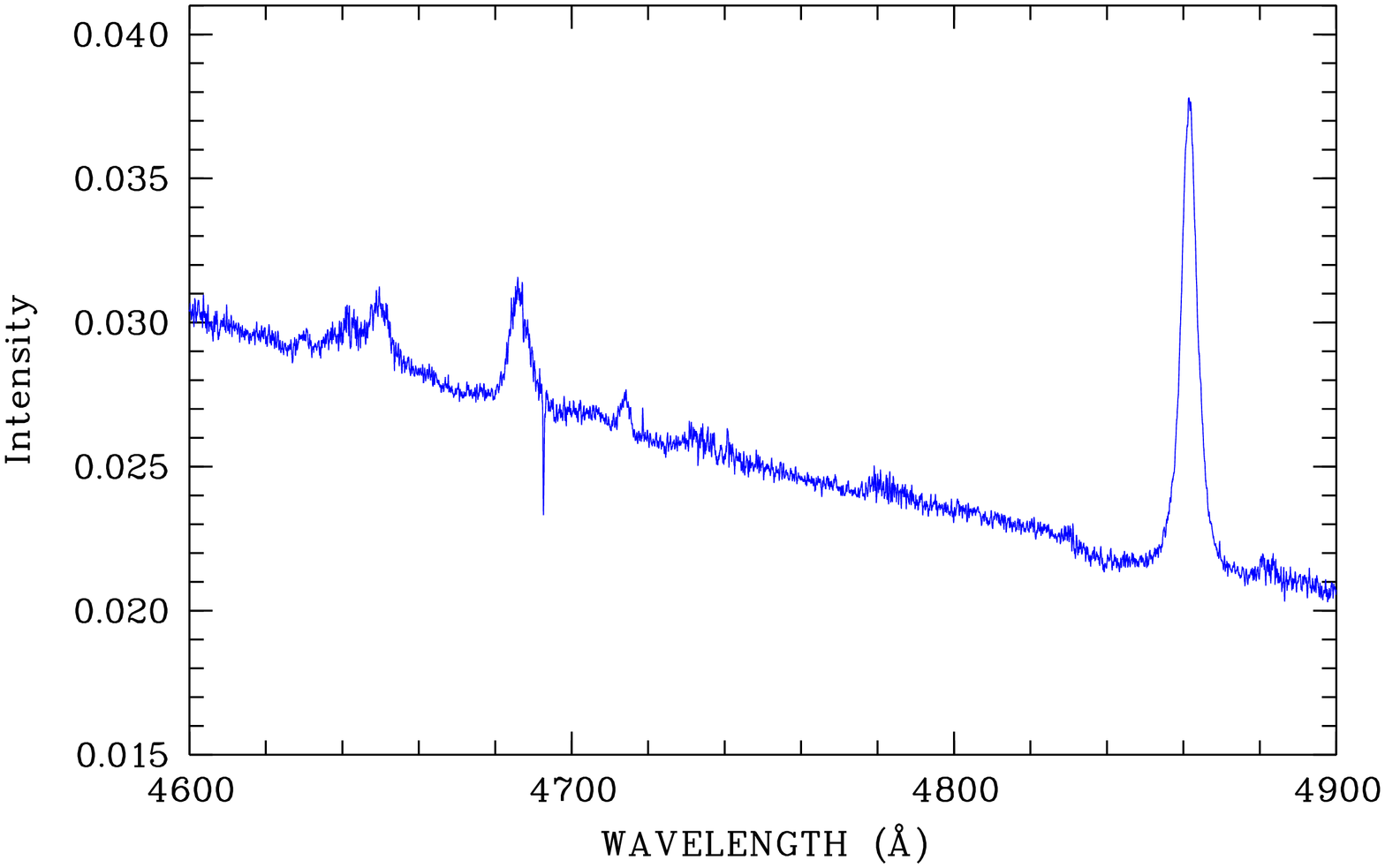}}
\caption{ANU/WiFeS low-resolution and SALT/HRS high-resolution identification spectra of \srgt. The SALT spectrum only has arbitrary uncalibrated intensity units. There is high excitation N III 4640\AA\, Bowen fluorescence and the He II 4540\AA\, and 4686\AA \,lines.\label{f:ids}}
\end{figure}

\subsection{Initial follow-up spectroscopy with WiFeS and SALT}
The likely \gai counterpart was selected for follow-up spectroscopy with the 2.3m telescope of the Australian National University, which is equipped with the Wide Field Spectrograph \citep[WiFeS;][]{Dopita2010Ap&SS.327..245D} and with the 10\,m class Southern African Large Telescope \cite[SALT; ][]{Buckley2006SPIE.6267E..0ZB}), which is equipped with the High Resolution Spectrograph \citep[HRS;][]{Crause+2014}. WiFeS is a double-beam, image-slicing, integral-field spectrograph. Observations were performed on 2020 October 28.164 with the R3000 grating and an exposure of 300\,s. As the red arm of the instrument was not operating, only the blue wavelength range between 3500 and 5500\,\AA\, is available. The WiFeS spectrum was reduced using the PyWiFeS reduction pipeline \citep{childress+14}, which produced three-dimensional data. We then extracted background-subtracted spectra from the slitlets that provided significant flux.

HRS is a dual-beam, fibre-fed èchelle spectrograph, covering 3800$-$8900\,\AA. A 1200\,s HRS observation was taken on 2020 October 31.9672 in low-resolution (LR; R $\sim$15,000) mode. Initial reduction of the HRS spectrum was achieved using the $\textit{PySALT}$ package \citep{Crawford2010SPIE.7737E..25C}, which includes overscan correction, bias subtraction, and gain correction. The spectrum was extracted using the HRS pipeline, based on MIDAS routines described in \cite{2016MNRAS.459.3068K}. The ANU/WiFeS and SALT/HRS spectra are displayed in Fig.~\ref{f:ids}. 

The object has a very blue continuum that increases to the short-wavelength cutoff at 3500\,\AA. The WiFeS spectrum reveals broad Balmer absorption lines whose centers are filled with intense emission lines. In addition to H-Balmer emission lines, the object displays He-emission lines, both neutral and ionized. Through these early spectra, the object is robustly identified as a CV. None of the spectra shows any sign of the donor star.

\begin{figure}
\resizebox{\hsize}{!}{\includegraphics{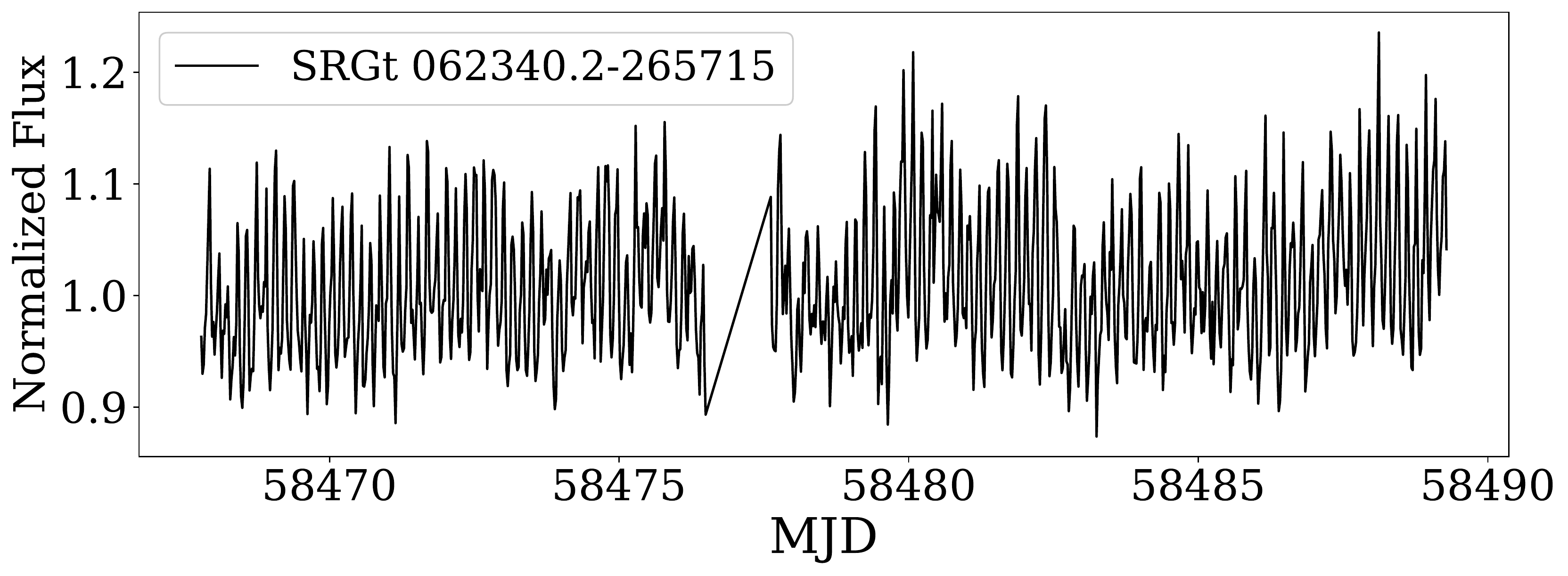}}
    \caption{\tes light curve for the SRGt 062340.2-265715. The normalized light curve was obtained by dividing the background-corrected source flux by its median value of 513.1 electrons \rat.}
    \label{f:lc}
\end{figure}

The typical full width at half maximum (FWHM) of the absorption lines corresponds to 3000\,\kmps, and those of the Balmer emission lines correspond to 300-380\,\kmps. The ionized helium line at 4686\,\AA\ is detected, but appears to be rather weak. It has an equivalent width of $\sim -0.9$\,\AA, an FWHM of $\sim$6\,\AA \,($\sim$380\,\kmps), and the flux ratio to the Balmer emission line H$\beta$ is $F(\ion{He}{ii})/F(H\beta) \simeq 0.26$.  H$\beta$ has an equivalent width of $\sim-4.2$\,\AA\, and an FWHM of $\sim$5.5\,\AA \, ($\simeq$340 \kmps). The HRS spectrum of the region around \ion{He}{ii}\,4686 also uncovers the Bowen blend of \ion{N}{iii} lines. This feature is often encountered in bright low-mass X-ray binaries \citep[LMXBs, e.g.,][]{cornelisse+08}, but also in the two types of magnetic CVs, the polars and the intermediate polars \citep[e.g.,][]{schachter+91,harlaftis_horne99}, indicating a strong UV/EUV source. 

The H$\beta$ emission line has a centroid of 4861.54\,\AA\ in the SALT spectrum, whereas it was found at 4862.59\,\AA\ in the WiFeS spectrum while being similarly broad (5.2\,\AA\ vs.~5.5\,\AA). The difference in wavelength corresponds to a velocity difference of 65\,\kmps , illustrating the feasibility of a radial velocity study.

\subsection{\tes}

We analyzed data from the \textit{Transiting Exoplanet Survey Satellite} \citep[\tes,][]{TESS2014SPIE}. \tes, using four CCD cameras, obtains continuous optical images of a rectangular field of $24^\circ \times 90^\circ$ for 27.4\,days at a short cadence (30\,minutes to 20\,seconds).

The source was observed in sector 6, which was observed at a 30-minute cadence between MJD 58467.79 to MJD 58489.54 (2018 December 15 to 2019 January 06). The source  was also observed in sector 33 at a faster 2-minute cadence  between MJD 59201.24 (2020 December 18) and MJD 59227.07  (2021 January 13).

We began the analysis with the data obtained in sector 6 (30\,min cadence) and used the Python package Lightkurve (v1) \citep{lcsoftware2018} to create the light curves from the full-frame images (FFI). From the FFI from which we extracted our light curves, we selected pixels around the coordinates of the target, and background or empty pixels with no known Gaia sources as our background and with a median flux that is lower than a given threshold.  We also used the \tes light curve from the MIT Quick-Look Pipeline \citep[QLP,][]{HuangQLP2020RNAAS...4..204H} and compared them to our light curves from the full-frame images using the median background-removal method. The results from the two programs agreed well. The final extracted light curve (30 min cadence), normalized to the median value of 513.1 electrons \rat, is displayed in Fig.~\ref{f:lc}. The light curve is flat to first order, but displays an apparent flickering that gives the impression of a possible periodic behavior. We thus performed a periodogram analysis to search for any underlying periodicity. 

\begin{figure}
\resizebox{\hsize}{!}{\includegraphics{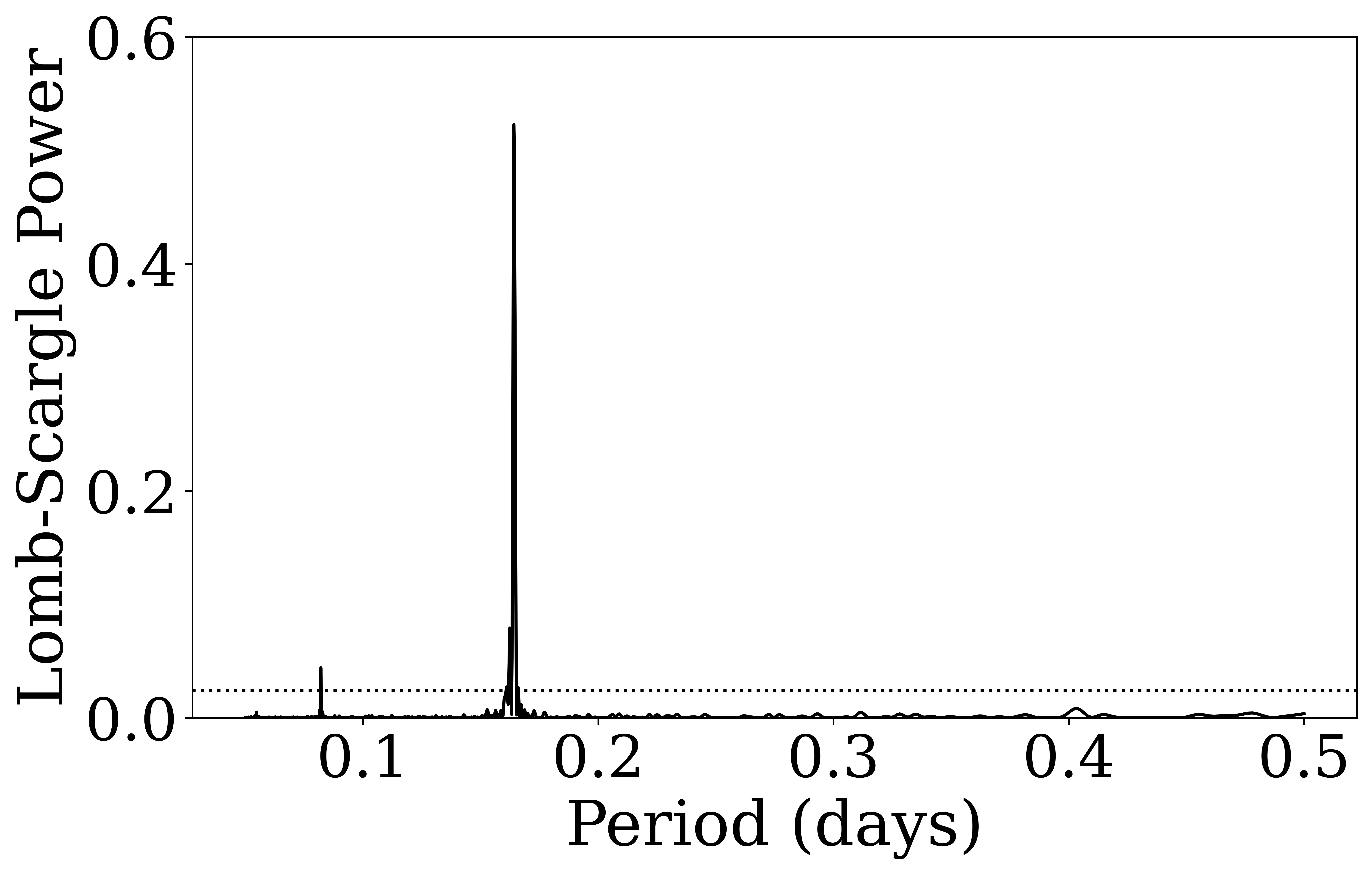}}
\caption{Lomb-Scargle periodogram for \srgt, showing a clear isolated peak around 3.941 hours. The dotted line shows the periodogram level corresponding to a maximum peak false-alarm probability of $1\%$ using the bootstrap method that simulates data at the same observation times to approximate the true distribution of peak maxima for the case without a periodic signal. The bootstrap method is performed with periods between 0.1 and 0.5 days and after normalizing by the residuals of the data around a constant reference model.
}
    \label{f:pgram}
\end{figure}

We performed period searches using the Lomb-Scargle \citep[LS,][]{lomb76,scargle82} 
and the phase dispersion minimization technique \citep[PDM,][]{Stellingwerf1978PDM} as implemented in the Astropy and the PyAstronomy packages \citep{astropy:2018,pyastronomypackage}. The LS-periodogram was then normalized by the residuals of the data around a constant reference model. The periodogram (Fig.~\ref{f:pgram}) clearly shows an isolated peak at a period of $3.941\pm0.010$\,h, the error being estimated from the width of the peak ($\sigma\sim 0.01$\,h). We also calculated a false-alarm probability of $3.9 \times 10 ^{-158}$ using the method described in \cite{Baluev}. 

In the PDM method, the light curve is divided into different phase bins, and the cost function, $\Theta =s^2/\sigma^2$, is minimized to choose the best period. Here $s$ is the phase bin variance and $\sigma$ is the total data variance. We divided the data into ten equidistant bins to calculate $\Theta$ and also found a period of 3.941\,hours, which is fully consistent with the results obtained with the Lomb-Scargle periodogram.

 \begin{figure}
\resizebox{\hsize}{!}{\includegraphics{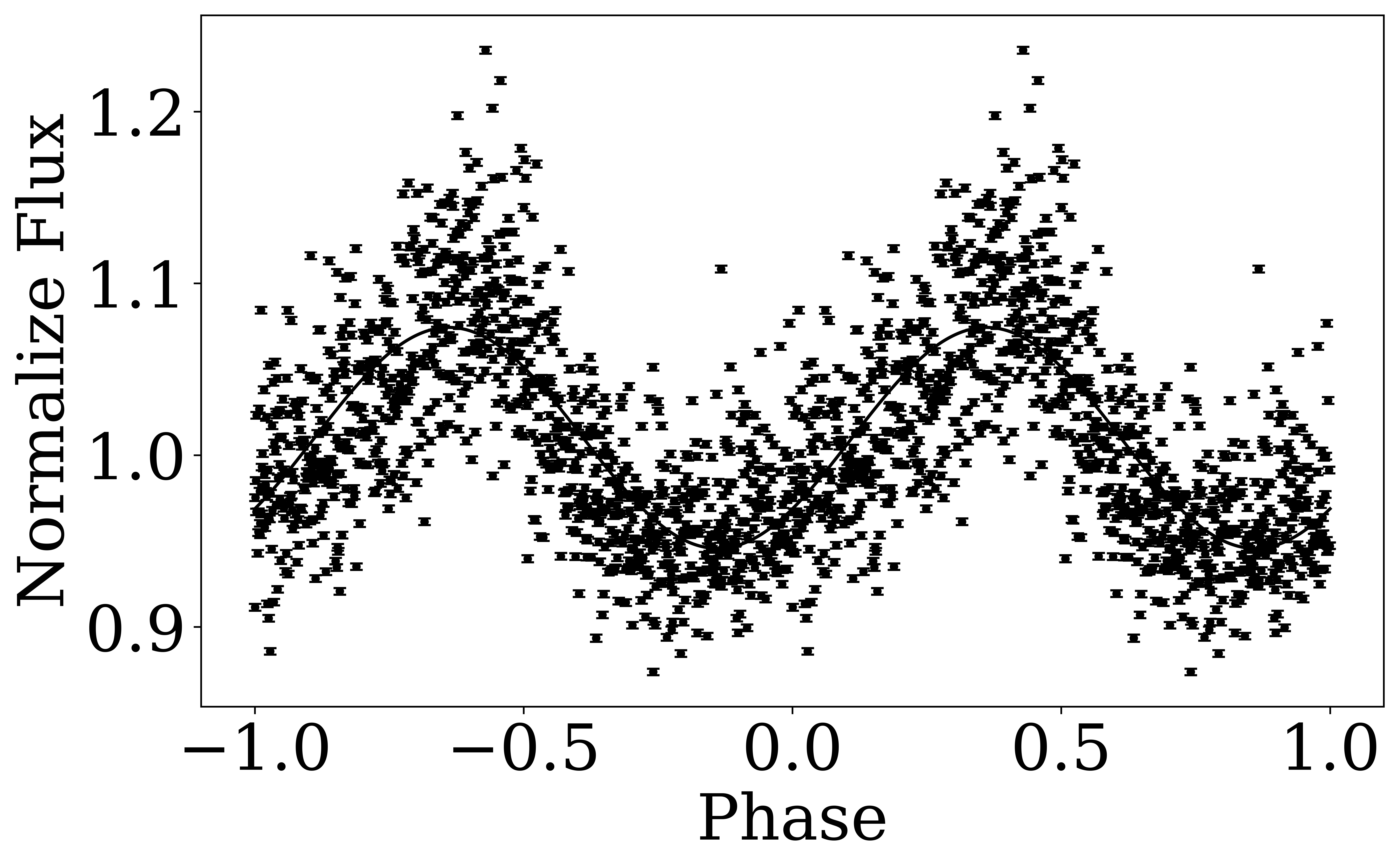}}
    \caption{\tes light curve for SRGt 062340.2-265715, folded on the 3.941-hour period. Data are shown with their error bars. The solid lines represent the best-fit sinusoid. Phase zero was chosen arbitrarily.}
    \label{f:folded}
\end{figure}

The main result of the analysis of the 30-minute cadence data, the presence of a periodicity at about 3.9\,hours, was reported very briefly already in ATEL 14222 by one of us \citep{pichardo20}. The more thorough analysis presented here confirms and refines the initial result. The light curve folded at the period of 3.941\,hours is shown in Fig.~\ref{f:folded}. It is slightly skewed, with a more gradual increase and a steeper decrease. There is significant scatter at any given phase, which indicates further variability on shorter timescales. This could not be resolved with the low-cadence \tes data of sector 6. 

\begin{figure}
\resizebox{\hsize}{!}{\includegraphics{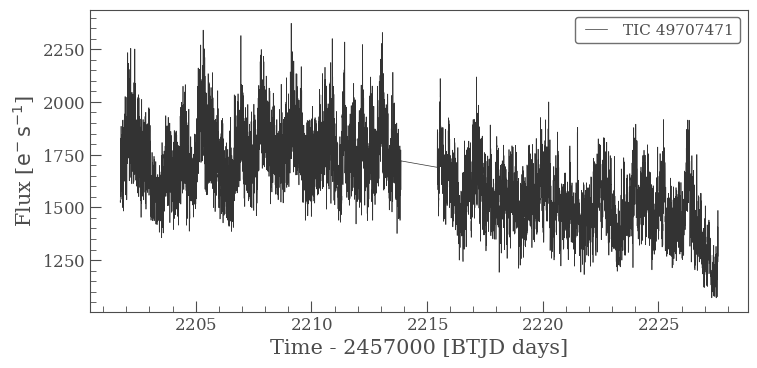}}
\caption{\tes light curve for \srgt obtained in sector 33 with a 2 min cadence.}
\label{f:t2m}
\end{figure}

\begin{figure}
\resizebox{\hsize}{!}{\includegraphics{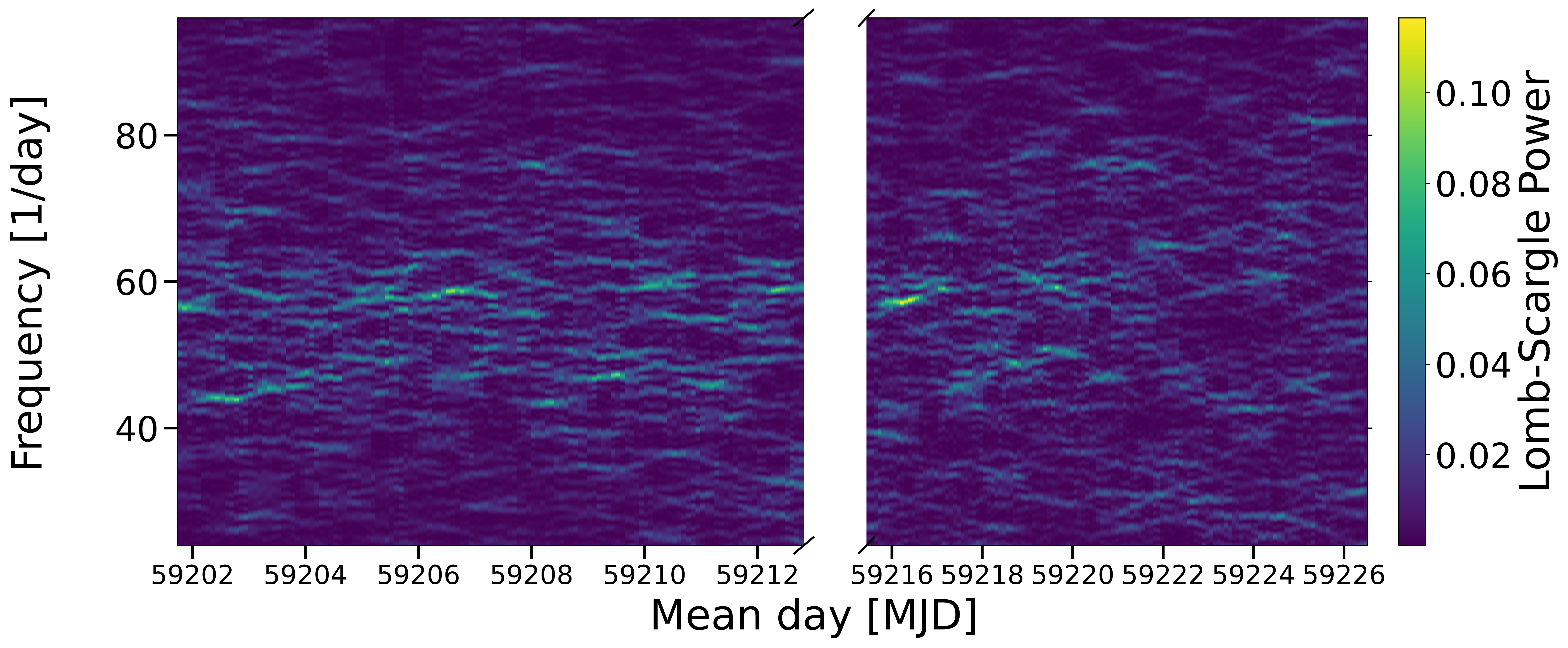}}
\caption{Moving-window periodogram of the \tes 2-minute cadence data for a sliding window of one-day length and time steps of two hours. The x-axis is the middle of the one-day window, and the color corresponds to the Lomb-Scargle power. The broken axis corresponds to the dates around the middle of the light curve that have some missing data. }
\label{f:ls2d}
\end{figure}

For the data obtained with a 2-minute cadence we used the Pre-search Data Conditioning Simple Aperture Photometry flux (PDCSAP), in which long-term trends have been removed using so-called cotrending basis vectors \citep[CBVs,][]{KeplerLCMethod2012}, produced by the \tes Science Processing Operations Center \citep[SPOC; ][]{TESSSPOC2016SPIE}. The light curve of data from sector 33 is shown in the original time sequence in Fig.~\ref{f:t2m}. On this occasion, the source appeared fundamentally different compared to the data obtained in sector 6. At the beginning of the observations, the source was about 1.4 mag brighter than two years before, and at the end of this one-month interval only by about 0.9 mag. A Lomb-Scargle period search for the whole dataset from sector 33 no longer showed the 3.9 h modulation. We then selected shorter time intervals of one  or two days length at the beginning, center, and end of the data train and searched for periodic behavior between 10 minutes and 5 hours. This revealed some power at 25 min. To study the stability of this period, we calculated a dynamic periodogram using a time slice with a length of one day that was moved forward with a step size of 2 hours. For each slice, an LS periodogram was computed, and the results were arranged as the two-dimensional periodogram (LS power as a function of time) that we show in Fig.~\ref{f:ls2d}. 

\begin{figure}
\resizebox{\hsize}{!}{\includegraphics{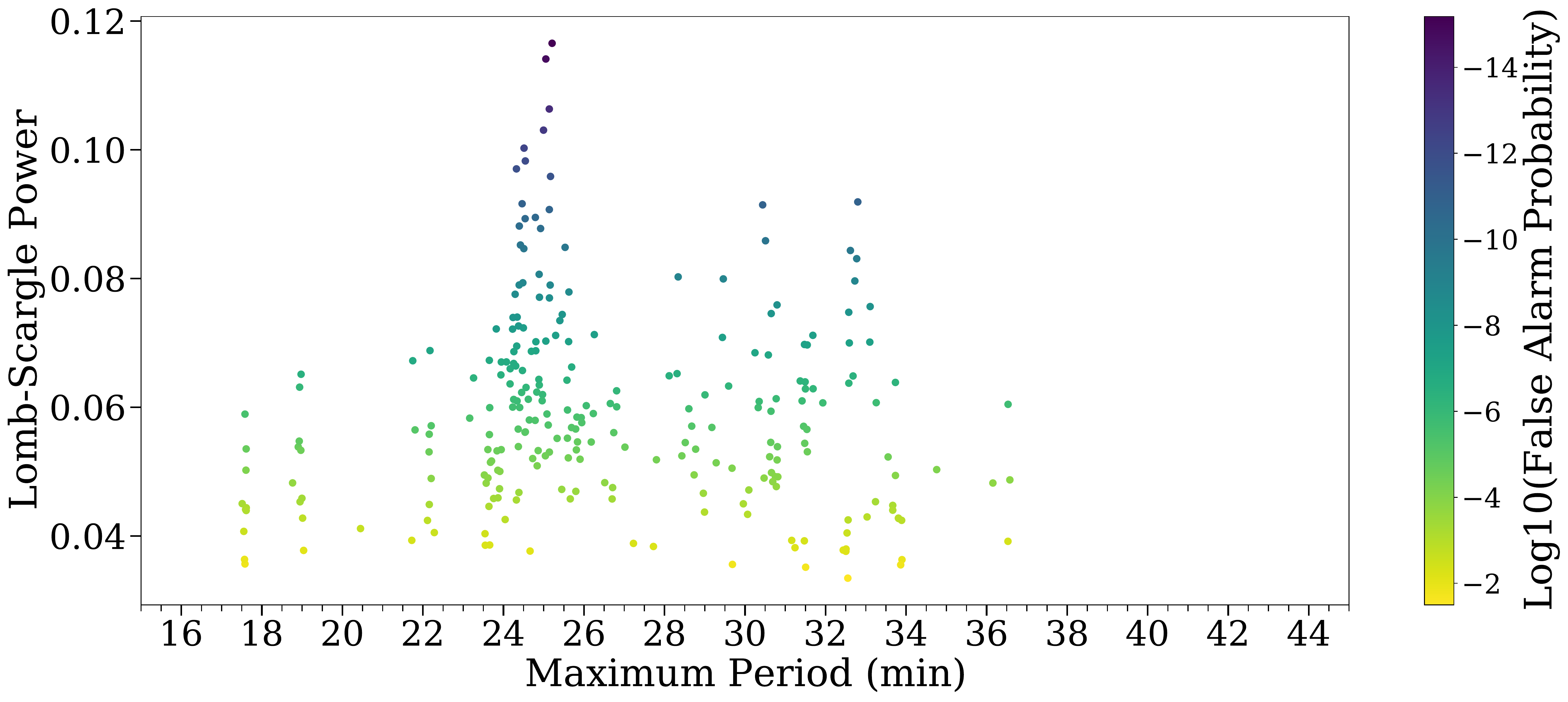}}
\caption{For each one-day window we plot the maximum amplitude of the periodogram and its corresponding period. The color corresponds to the $\log_{10}$ of the FAP.}
\label{f:ls2dfap}
\end{figure}

\begin{figure}
\resizebox{\hsize}{!}{\includegraphics{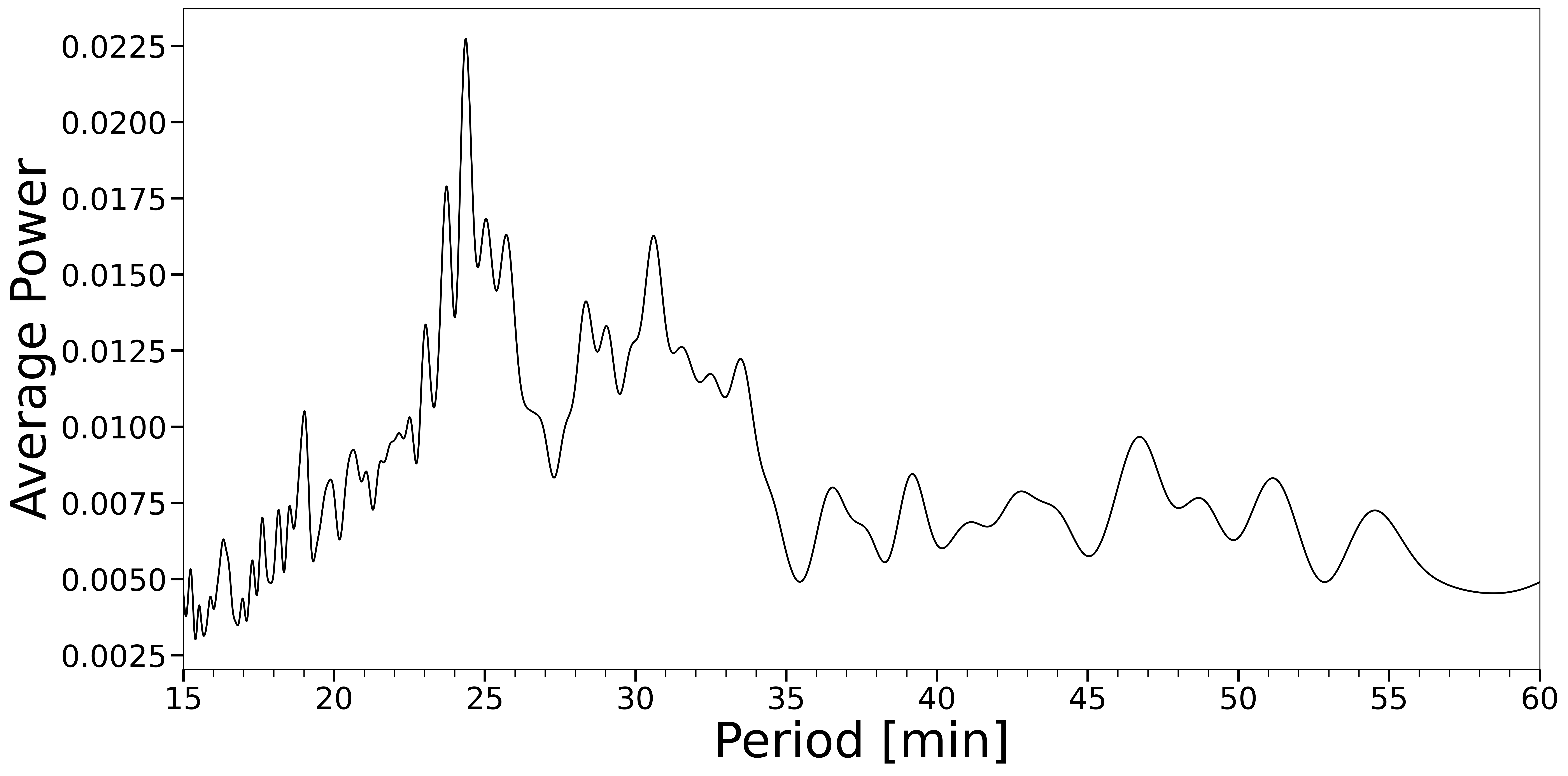}}
\caption{Mean periodogram of sector 33 calculated by averaging all one-day periodograms of Fig.~\ref{f:ls2d}.}
\label{f:ls2dmean}
\end{figure}

Complementary views on the complex timing behavior of the source are given in Figs.~\ref{f:ls2dfap} and \ref{f:ls2dmean}, which show the false-alarm probabilities of the highest peak in each of the one-day periodograms and the mean periodogram of all the individual one-day periodograms. We found significant power centered around 25 min (frequency 57.6 d$^{-1}$), but measured significant frequencies that jumped between 45 and 65 cycles per day on timescales of about one day. No significant period was found during the last quarter of the data set. The mean periodogram has a maximum power at 24.37 min, and the most significant period in a one-day time slice is observed at 25.201 min with a false-alarm probability of $6.58 \times 10^{-16}$. 

\begin{figure}[t]
\resizebox{\hsize}{!}{\includegraphics{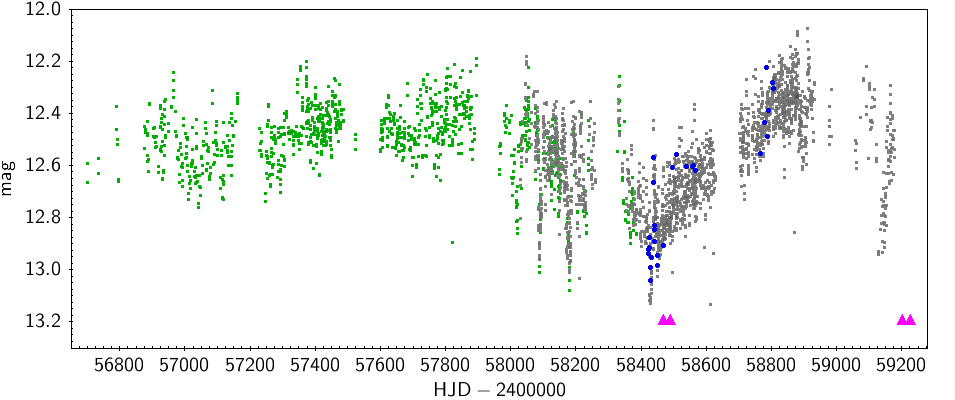}}
\caption{ASASS-SN and ZTF photometric data for \srgt (green symbols show ASAS-SN $V$, gray symbols show ASAS-SN $g$, and blue symbols represent ZTF $g$). The magenta triangles mark \tes observations with low and high cadence (the first and second pair identify \tes sectors 6 and 33, respectively).}
\label{f:alcs}
\end{figure}

\subsection{Archival photometry from CRTS, ZTF, ASAS-SN, and ATLAS}
Archival photometric information of the new transient is available from the Catalina Real-Time Transient survey \citep[CRTS,][196 epochs]{drake+09}, the Zwicky Transient Facility \citep[ZTF,][55 epochs]{masci+19}, the All-Sky Automated Survey for Supernovae \citep[ASAS-SN,][3350 epochs]{shappee+14,kochanek+17}, and the Asteroid Terrestrial-impact Last Alert System \citep[ATLAS,][34 epochs]{tonry+18,heinze+18}. The observations of these surveys cover the time intervals 2005 August 28 to 2010 April 19, 2018 November 1 to 2019 November 18, 2014 February 11 to 2020 November 24, and 2021 February 16 to 2021 March 16, respectively, and part of them are shown in Fig.~\ref{f:alcs} (omitting the CRTS and ATLAS data). 
ZTF data were taken in the $g$ and $r$ filters with only $g$-band data shown, and ASAS-SN data through $g$ and $V$ filters. All these data show extended phases (years) with little variability around a mean magnitude of about 12.5, with superposed scatter with an amplitude of 0.1 mag followed by dimmed phases with a larger short-term variability amplitude. These excursions toward lower brightness were observed twice, one at the end of the CRTS data train at MJD 56350, and the other was covered by both ZTF (partially) and ASAS-SN at MJD 58400. The minimum brightness of these fainter states is at around 13 mag. \tes observations in sector 6 revealing the 3.9 hour periodicity were obtained at the pronounced minimum. No automated survey data (e.g., ASAS-SN or ATLAS) are available at the time of writing when \tes observed the source again in sector 33.

We searched for periodic behavior of the source in the ASAS-SN data. We selected $g-$band data obtained between HJD 2458455 and 2458866. During this 400-day interval with 1072 data points, the transient recovered from its faint state at $g=12.95$ to a bright state at $g=12.3$ with a constant gradient. The brightness displayed a scatter with a mean amplitude of 0.085 mag around the trend, while the mean magnitude error was only 0.01 mag. We used the LS period search method as implemented in ESO-MIDAS and found a period $3.9142 \pm0.0011$ hours, which is consistent with the value found in the \tes data at 2.5$\sigma$.

\begin{figure}[t]
\resizebox{\hsize}{!}{\includegraphics{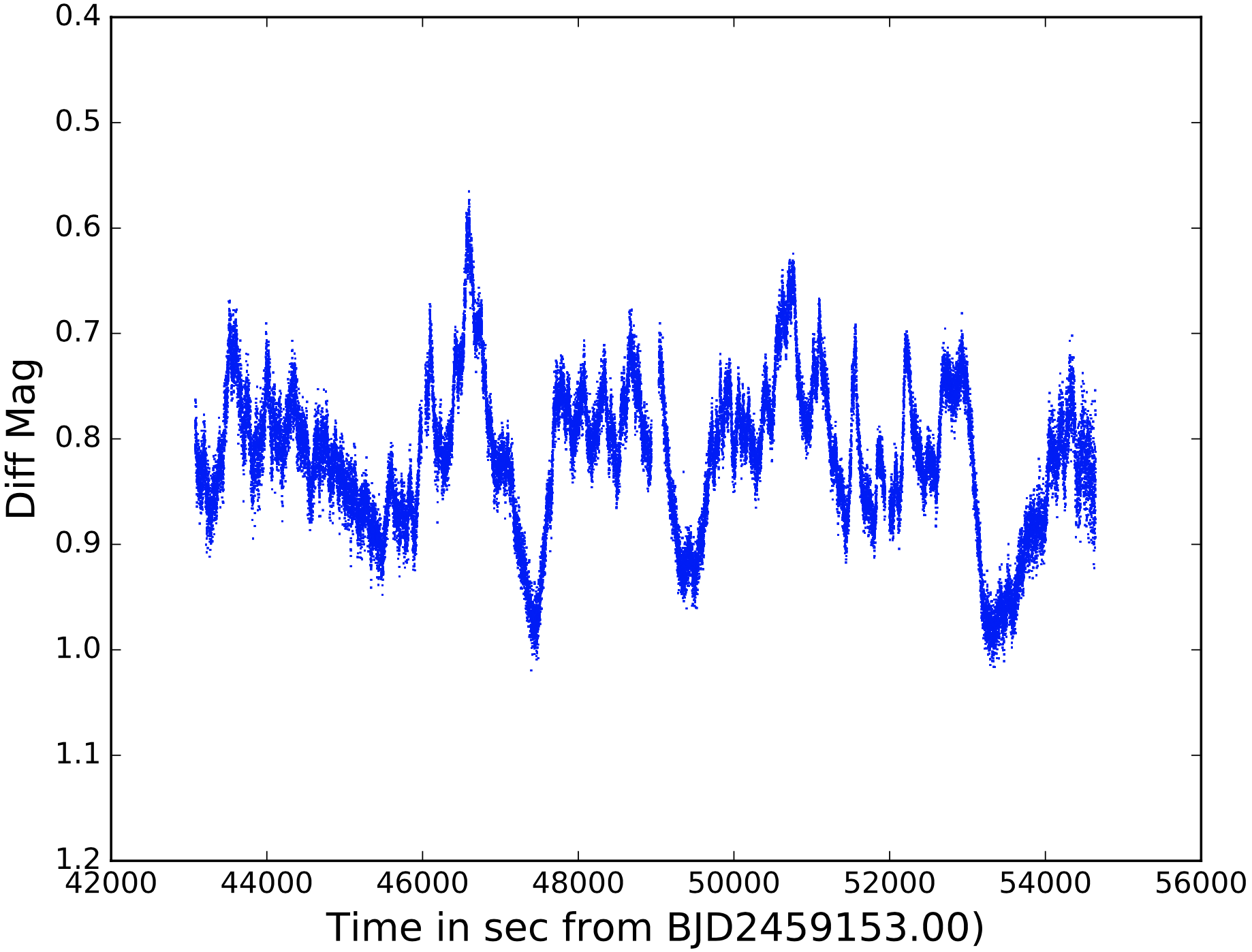}}
\caption{Differential white-light optical light curve, spanning 3.2 h, of \srgt  obtained with \emph{SAAO/SHOC} at 0.3 s cadence. Time is in seconds since 00:00 UT on 1 November 2020.}
\label{f:shoc-lc}
\end{figure}

\begin{figure}
\resizebox{\hsize}{!}{\includegraphics{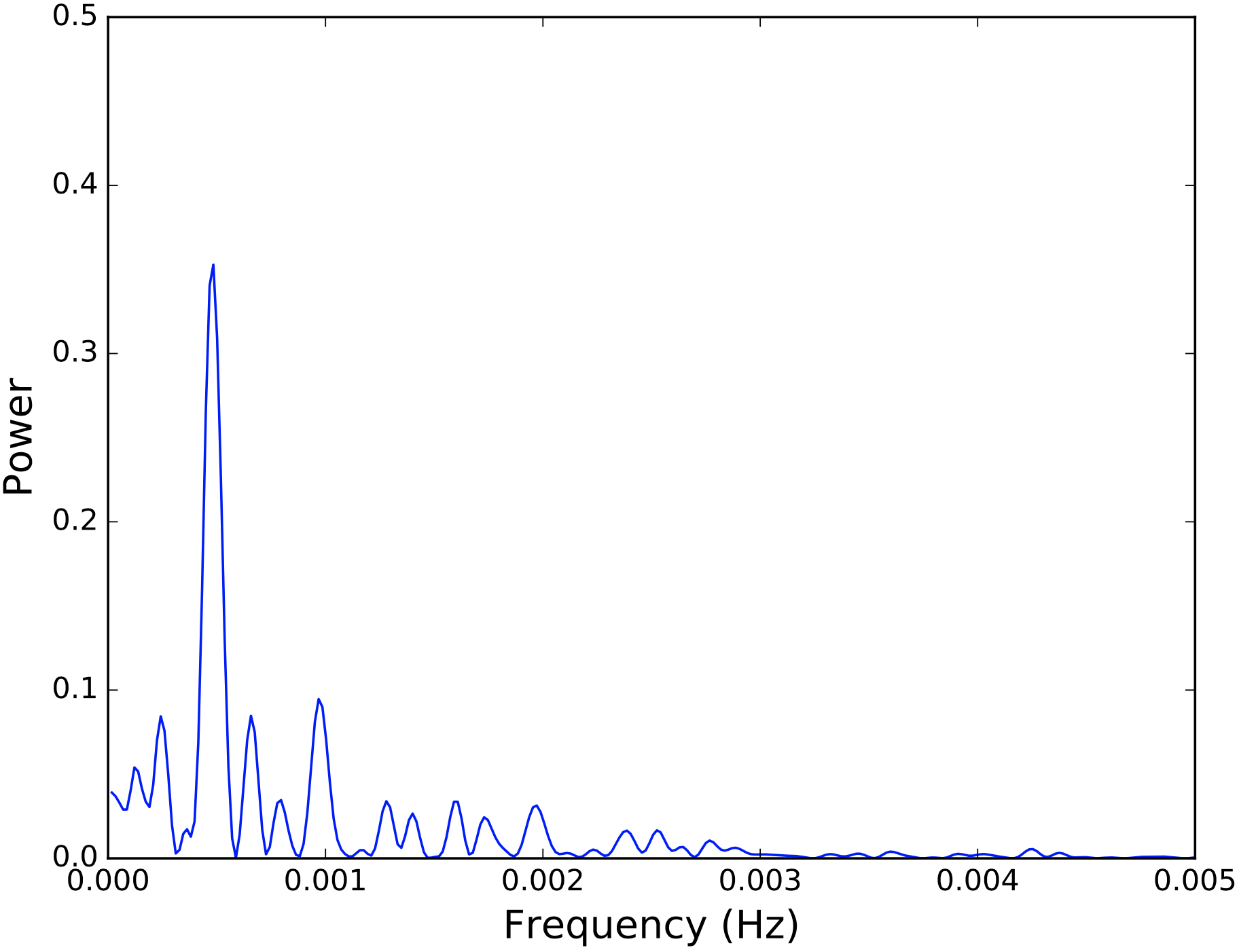}}
\caption{Lomb-Scargle power spectrum of the light curve of \srgt (shown in Fig~\ref{f:shoc-lc}) indicating a possible $\sim$2100\,s periodicity.}
\label{f:shoc-ls}
\end{figure}

\subsection{SAAO high-speed photometry}
Following its spectral identification (see subsection 2.4), we performed 3.2 h of high-speed photometry with the South African Astronomical Observatory 1.0 m telescope on 2020 November 1, beginning at 23:57:30 UTC. The Sutherland High speed Optical Camera \citep[SHOC;][]{Coppejans2013PASP..125..976C} CCD camera, which uses an Andor iXon888 frame transfer EM-CCD frame camera (1024 $\times$ 1024 pixels), was used. The observations were made without a filter (i.e., `white light') with an exposure time of 0.3~s. 

The CCD images were reduced using the $\it{TEAPhot}$ photometry reduction package \citep{2019A&A...629A..21B} and included subtraction of median bias and flat-field correction using median-combined frames from exposures of the twilight sky. $\it{TEAPhot}$ employs the method of adaptive elliptical aperture photometry, which was used to create calibrated science images and a differential light curve using the bright (V = 12.1) reference star, UCAC2 20781392. The differential light curve is shown in Fig.~\ref{f:shoc-lc}. 

The light curve shows strong flickering that is typical of CVs and is reminiscent of some intermediate polars that exhibit spin-related periodic modulations. We therefore subjected the light curve to a Lomb-Scargle period analysis, probing frequencies as high as the Nyquist limit (0.6 s). Except for a strong peak at 0.480 mHz, there were no other significant peaks.  The result is shown in Fig.~\ref{f:shoc-ls}. The strong peak corresponds to a period of $2080 \pm 100$\,s (34.7 min). 

\begin{figure}[t]
\resizebox{\hsize}{!}{\includegraphics{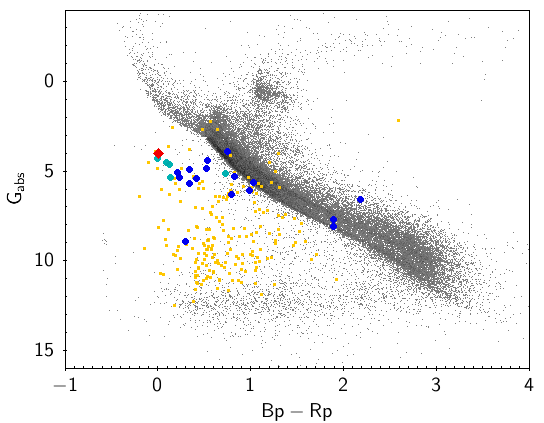}}
\caption{Color-magnitude diagram for eRASS2 sources.
Each gray dot is an eRASS2 source that is matched (within 8 arcsec) with a Gaia source within 500 pc (distances by Bailer-Jones et al.~2018). In dark yellow we plot all CVs from the final Ritter \& Kolb catalogue, in green the nonmagnetic high-accretion rate UX UMas therein, and in blue the intermediate polars from Mukai's compilation. The new CV \srgt is identified by the red symbol.}
\label{f:hrd}
\end{figure}

\section{Results and discussion\label{s:res}}

We have analyzed space- and ground-based data of \srgt, an object that was initially detected as an X-ray transient with the two X-ray instruments on board SRG. Follow-up data, in particular the initial spectroscopy, uniquely identify the object as a CV and thus confirm Denisenko's previous classification. At $G= 12.5$\,mag the object is clearly one of the brighter examples of its class. An important question is to which, if any, CV family the object belongs.

The obtained spectra and the location of the object in the color-magnitude diagram (see Fig.~\ref{f:hrd}) all strongly suggest that it belongs to the novalike subclass. No dwarf nova outburst was recorded in the multi-year monitoring observations by CRTS. 

The great majority of novalikes are found above the CV period gap. It is thus tempting to associate the 3.9-hour periodicity found in ASAS-SN and \tes data (sector 6) with the orbital period of the binary. The disappearance of this period in \tes sector 33 data is puzzling. It might be premature to associate the 3.9-hour periodicity with the orbital period , and alternatively, the photometric phenomenology might be completely changed in the much higher accretion state of \tes sector 33 data. Time-resolved spectroscopy is needed to uniquely identify the orbital period by tracing the absorption lines and the emission cores through an orbital cycle. The blue optical continuum and the absorption lines are thought to originate from an optically thick accretion disk. The narrow emission lines might be resolved into two components, one to be associated with the cooler outer parts of the disk and the other originating from the irradiated hemisphere of the donor star \cite[see, e.g.,][for templates]{beuermann+thomas90,hernandez+17}, each with its own radial velocity curve. 

The production sites of X-rays in nonmagnetic novalikes are debated. Originally, X-rays were assumed to originate from the boundary layer (BL) between the disk and the white dwarf with a blackbody-like spectral shape, but the predicted blackbodies were not found \citep{mauche_mukai02}. The BL is nevertheless mostly regarded as the X-ray production site \citep{mukai17}, but other locations such as shocked circumstellar material, the polar caps, a central corona, a partially obscured boundary layer, or advective hot flows are discussed \citep{zemko+14, dobrotka+17, balman20}. The already observed high degree of X-ray variability on long and short timescales, that is,~between eRASS surveys and between erodays, together with dedicated spectral observations may shed new light on this old question. If it belongs to the nonmagnetic novalikes, \srgt\ likely does not belong to the VY Scl subclass because it does not show the typical anti-dwarf nova dimmings by $\sim$4 mag. 

Accretion in the intermediate polars may occur through streams and disks. The disks may also be large and extend up to the Roche lobe of the accreting white dwarf \citep{hellier+91}. The freely spinning white dwarf in an IP typically leads to very rich power spectra, in particular at X-ray wavelengths, in which orbital, spin, and different types of beat periods can be traced \citep[dependent on the orbital inclination and the accretion mode, respectively, see, e.g.,][]{norton+96}. The high degree of X-ray variability through self-eclipses and foreshortening of the accretion spots and internal absorption by accretion curtains is common to all IPs. The observed large X-ray variability within eRASS2 and between the two SRG surveys may thus argue for an IP classification. Moreover, the SAAO-discovered period of 35 min  and the \tes-discovered period at 25 min might support this view because if these periods are associated with the spin of the white dwarf, they are typical of IPs. In particular, the \tes-discovered periods at 25 min indicate an IP nature because many IPs are observed to cluster around period ratios of $P_{\rm spin} / P_{\rm orb} = 0.1$ \citep[][their figure 1]{norton+04}, as implied for \srgt.  To firmly classify the object as an IP, however, the detection of a stable spin period is required, which is lacking. The appearance and disappearance of periods at around half an hour is perhaps difficult to explain in either classification.

Optical power spectra of IPs are, however, affected by reprocessed radiation. Better insight is possible from a decent uninterrupted X-ray observation to uncover the intrinsic variability of the source on short timescales. The existing X-ray observations of \srgt with SRG are insufficient to detect a $\sim$2000 s period. This will require an extensive future observation (e.g.,~with XMM-Newton).

The X-ray spectrum of \srgt is not too strongly absorbed and can be modeled with a single purely thermal emission component. No separate soft X-ray component was found, although the appearance of the Bowen blend appears to imply the existence of such a component. The eRASS2 bolometric X-ray flux implies an accretion luminosity of $L_{\rm X, bol} = 4.8 \times 10^{32}$\,erg \rat , which is at the low end for an IP and at the high end for an UX UMa system.

\begin{figure}[t]
\resizebox{\hsize}{!}{\includegraphics{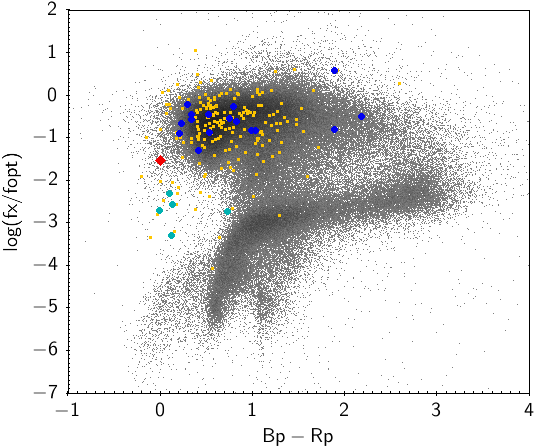}}
\caption{X-ray to optical color-color diagram. Gray background objects are all eRASS2 sources that match Gaia within 8 arcsec. Symbol colors are otherwise the same as in Fig.~\ref{f:hrd}.}
\label{f:ccd}
\end{figure}

In addition to the color-magnitude diagram of Fig.~\ref{f:hrd}, it is instructive to locate \srgt in an X-ray to optical color-color diagram based on \ero and \gai, which is shown in  Fig.~\ref{f:ccd}. The diagrams were constructed in the following manner. The collaboration-internal preliminary eRASS2 catalog was matched to within the 8 arcsec FHWM of the point spread functions of the telescope modules of \ero\ with \gai\ DR2. For the matching objects, distances were  determined following \cite{bailer-jones+21}. We then used the final edition of the Ritter \& Kolb catalog \citep{ritter_kolb03} to locate the CVs. The optically bright novalike CVs of the UX UMa type are identified with a different color in the diagrams. We also used the compilation of Koji Mukai\footnote{https://asd.gsfc.nasa.gov/Koji.Mukai/iphome/iphome.html} to highlight the IPs that were identified in eRASS. The color $Bp - Rp$ was taken directly from \gai data. The X-ray to optical color was computed as $\log(F_{\rm X} (0.6-2.3 {\rm keV}) + {\rm phot\_g\_mean\_mag}/2.5 + 4.86$, and the chosen energy band was the standard band 2 in the current version of the catalog pipeline \citep[pipeline version 946][submitted to A\&A]{brunner+21}.

Our Fig.~\ref{f:hrd} is to be compared with Fig.~2 of \cite{abril+20}, which locates the CV sub-families in the \gai-DR2 HRD. Whether the floor of objects in Fig.~\ref{f:hrd} at $G=13$\,mag indicates real associations or is due to random associations needs to be determined. Most CVs are located between the white dwarf sequence and the main sequence. Their X-ray luminosities are orders of magnitude higher than those of coronally active stars on the main sequence. The novalike CVs, in particular the magnetic IPs, overlap in the color-magnitude diagram with coronally active stars, but they can be discerned through their much higher X-ray luminosity. 

The color-color diagram of Fig.~\ref{f:ccd} has two main structures. The L-shaped sequence beginning at $(0.6,-5.5)$ with the knee at $(1,-3)$ delineates active stars, whereas the cloud around $(1,-0.5)$ is the domain of the active galactic nuclei. Most of the CVs cannot be distinguished from the numerous active galactic nuclei in this diagnostic diagram as long as no astrometric information can be used for a further distinction of galactic and extragalactic objects \citep[see also][their Fig.~11]{comparat+20}.

The transient CV \srgt is located in very sparsely populated areas in both diagrams. It optically belongs to the most luminous known X-ray detected CVs. It is bluer than most nonmagnetic UX UMa novalike CVs and bluer than all IPs, although it is not too much separated from the extreme representatives of these two classes. On the other hand, its X-ray to optical color is between that of the UX UMas and the IPs. For an IP, \srgt is X-ray underluminous by about 1\,dex; for a nonmagnetic novalike, it is also overluminous by about 1\,dex. 

At a \gai magnitude of 12.5\,mag, \srgt, if identified as an IP, would be the IP with the brightest apparent magnitude in eRASS2 by far. The next brightest IP is TV\,Col at $\mathrm{phot\_g\_mean\_mag} = 13.9$\,mag. For comparison, the first soft X-ray discovered IP in the \ros all-sky survey, PQ Gem, was observed at  $\mathrm{phot\_g\_mean\_mag} = 14.1$\,mag.

Although only a few objects are found in the vicinity of \srgt in the diagnostic diagrams of Figs.~\ref{f:hrd} and~\ref{f:ccd}, several candidates are located around our new discovery that will be identified in the next years with the large-scale identification programs of serendipituous \ero sources with the SDSS \citep{kollmeier+17}, for instance. The journey has just begun.

\begin{acknowledgements}
We thank the referee, John Thorstensen, for constructive criticism.

MPM would like to thank Tom Maccarone for useful discussions.

This work is based on data from \ero, the soft instrument aboard SRG, a joint Russian-German science mission supported by the Russian Space Agency (Roskosmos), in the interests of the Russian Academy of Sciences represented by its Space Research Institute (IKI), and the Deutsches Zentrum für Luft- und Raumfahrt (DLR). The SRG spacecraft was built by Lavochkin Association (NPOL) and its subcontractors, and is operated by NPOL with support from the Max Planck Institute for Extraterrestrial Physics (MPE).

The development and construction of the eROSITA X-ray instrument was led by MPE, with contributions from the Dr. Karl Remeis Observatory Bamberg \& ECAP (FAU Erlangen-Nuernberg), the University of Hamburg Observatory, the Leibniz Institute for Astrophysics Potsdam (AIP), and the Institute for Astronomy and Astrophysics of the University of Tübingen, with the support of DLR and the Max Planck Society. The Argelander Institute for Astronomy of the University of Bonn and the Ludwig Maximilians Universität Munich also participated in the science preparation for eROSITA.

The eROSITA data shown here were processed using the eSASS/NRTA software system developed by the German eROSITA consortium.

The {\it Mikhail Pavlinsky} ART-XC telescope is  the  hard X-ray instrument on board the \srg\ observatory, a flagship astrophysical project of the Russian Federal Space Program realized by the Russian Space Agency, in the interests of the Russian Academy of Sciences. ART-XC was developed by the Space Research Institute (IKI, Moscow) and the Russian Federal Nuclear Center -- All-Russian Scientific Research Institute for Experimental Physics (RFNC-VNIIEF, Sarov) with the participation of the  NASA's Marshall Space Flight Center (MSFC). The ART-XC team thanks the Russian Space Agency, Russian Academy of Sciences and State Corporation Rosatom for the support of the \srg\ project and ART-XC telescope. 

Some of the observations presented here were obtained with SALT under the transients followup programme 2018-2-LSP-001 (PI: DB), which is supported by Poland under grant no. MNiSW DIR/WK/2016/07.

DB and LT also acknowledge research support from the National Research Foundation.

We acknowledge the use of \tes High Level Science Products (HLSP) produced by the Quick-Look Pipeline (QLP) at the TESS Science Office at MIT, which are publicly available from the Mikulski Archive for Space Telescopes (MAST). Funding for the TESS mission is provided by NASA's Science Mission directorate.

This work has made use of data from the Asteroid Terrestrial-impact Last Alert System (ATLAS) project. The Asteroid Terrestrial-impact Last Alert System (ATLAS) project is primarily funded to search for near earth asteroids through NASA grants NN12AR55G, 80NSSC18K0284, and 80NSSC18K1575; byproducts of the NEO search include images and catalogs from the survey area. This work was partially funded by Kepler/K2 grant J1944/80NSSC19K0112 and HST GO-15889, and STFC grants ST/T000198/1 and ST/S006109/1. The ATLAS science products have been made possible through the contributions of the University of Hawaii Institute for Astronomy, the Queen’s University Belfast, the Space Telescope Science Institute, the South African Astronomical Observatory, and The Millennium Institute of Astrophysics (MAS), Chile.

This paper includes data collected with the TESS mission, obtained from the MAST data archive at the Space Telescope Science Institute (STScI). Funding for the TESS mission is provided by the NASA Explorer Program. STScI is operated by the Association of Universities for Research in Astronomy, Inc., under NASA contract NAS 5–26555.

\end{acknowledgements}

\bibliographystyle{aa}
\bibliography{srgt}

\end{document}